\documentclass[aps,prb,twocolumn]{revtex4}
\usepackage[dvips]{color}
\usepackage[normalem]{ulem}

\definecolor{g-blue}{rgb}{0.83,0.95,1}
\definecolor{g-yellow}{rgb}{1,1,0.7}
\definecolor{g-green}{rgb}{0.9,1,0.9}
\definecolor{green}{rgb}{0,0.6,0}
\definecolor{cyan}{rgb}{0,0.7,0.7}
\definecolor{black}{rgb}{0,0,0}
\definecolor{grey}{rgb}{0.4 ,0.4 ,0.4 }

\usepackage{bm}
\usepackage{amssymb}
\usepackage{amsfonts}
\usepackage{amsmath}
 \usepackage{graphicx}
  \usepackage{amsmath,bm,epsfig}
\def \ed {\end{document}}
\def\Fbox#1{\vskip1ex\hbox to 8.5cm{\hfil\fboxsep0.3cm\fbox{%
  \parbox{8.0cm}{#1}}\hfil}\vskip1ex\noindent}  

\newcommand{\Eq}[1]{Eq.\,(\ref{#1})}
\newcommand{\Eqs}[1]{Eqs.\,(\ref{#1})}
\newcommand{\Fig}[1]{Fig.\,\ref{#1}}

\def\be{\begin{equation}}\def\ee{\end{equation}}
\def\bea{\begin{eqnarray}}\def\eea{\end{eqnarray}}
\def\bse{\begin{subequations}}\def\ese{\end{subequations}}
\newcommand{\BE}[1] {\begin{equation}\label{#1}}
\newcommand{\BEA}[1]{\begin{eqnarray}\label{#1}}
\newcommand{\BSE}[1]{\begin{subequations}\label{#1}}

\let \= \equiv  \let\*\cdot \let\~\widetilde \let\^\widehat \let\-\overline

  \def\1{\bm1} 

\def\<{\left\langle}    \def\>{\right\rangle}
\def\({\left(}          \def\){\right)}
 \def \[ {\left [} \def \] {\right ]}

\newcommand{\B}[1]{{\bm{#1}}}
\newcommand{\C}[1]{{\mathcal{#1}}}    
\newcommand{\BC}[1]{\bm{\mathcal{#1}}}

\renewcommand{\sb}[1]{_{\text {#1}}}  
\def\Sb#1{_{\scriptscriptstyle\rm{#1}}}
\def\He4 {$^4$He~}

\begin{document}

\title{Dynamics of the vortex line density in superfluid counterflow turbulence}
\author{ D. Khomenko$^1$, V. S. L'vov$^1$, A. Pomyalov$^1$ and  I. Procaccia$^{1,2}$ }
\affiliation{$^1$The Department of Chemical Physics, The Weizmann Institute of Science,
Rehovot 76100, Israel\\$^2$Niels Bohr International Academy, University of Copenhagen
Blegdamsvej 17, DK-2100 Copenhagen, Denmark}

\begin{abstract}  Describing  superfluid turbulence at intermediate scales between the inter-vortex distance and the macroscale requires an acceptable equation of motion for the density of quantized vortex lines $\C L$. The  closure of such an equation for superfluid inhomogeneous flows requires additional inputs besides $\C L$ and the normal and superfluid velocity fields.   In this paper we offer a  minimal  closure using one additional anisotropy parameter $I_{l0}$. Using the example of counterflow superfluid turbulence we derive two coupled closure equations for the vortex line density and the anisotropy parameter $I_{l0}$ with an input of the normal and superfluid velocity fields. The various closure assumptions and the predictions of the resulting theory are tested against numerical simulations.
\end{abstract}

\maketitle

\section{Introduction}
\label{intro}

The experimental study of the statistical properties of mechanically excited $^4$He superfluid turbulence  was somewhat
stalled by a realization that these statistics, when observed on large spatial scales, do not
differ much from the statistics of the classical counterpart \cite{tabeling98}. These results could mean that the two-fluid model of superfluid turbulence may suffice to predict the statistical characteristics of superfluid turbulence. On the other hand, even earlier work on {\em counterflow} turbulence \cite{Vinen1957}, showed very definitely
that on scales that are intermediate between the inter-vortex distances $\ell$ and the outer scale $L$ the observed physics is influenced dramatically by vortex dynamics, vortex reconnections, Kelvin waves on the individual vortex lines, or in short, those effects that exist due to the quantization of vorticity in quantum fluids.  These findings underline the difference between ``co-flowing" situations (in which the normal and super components average flow is in the same direction) from ``counter-flowing" situations. Evidently, a fuller understanding of the statistical theory of ``counterflow" superfluid turbulence on these interesting scales calls for deriving an equation for the relevant characteristics of the tangle of quantized vortices which is ubiquitous in this type of turbulence. Indeed, the search for an equation of one of these characteristics, i.e. the density of vortex lines $\C L$, has been long, starting
with the seminal papers of Vinen from the nineteen fifties \cite{Vinen1957}. The context in which this
equation was studied was that of ``counterflow" turbulence in which $^4$He at temperatures below the $\lambda$-point is put in a channel with a temperature gradient, such that the mean normal velocity $\B V_n$ is directed from the hot to the cold end while the mean superfluid velocity $\B V_s$ is directed oppositely. The difference between these two mean velocities was denoted as $\B V_{\rm ns}$ and the equation that was proposed
by Vinen may be written as:
\begin{equation}\label{VinenEq}
\frac{d\C L}{dt}=\alpha C_1 |\B V\sb{ns}|\C L^{3/2}-\alpha C_2 \kappa \C L^2 \ ,
\end{equation}
where $C_1$ and $C_2$ are dimensionless coefficients, $\alpha$ is known as the ``mutual friction parameter" and $\kappa\approx  0.001  \rm{cm^2/s}$ is the quantum of circulation.
Remarkably, this equation which was not derived from first principles, and which was explicitly assumed to
apply to homogeneous isotropic situations, has been employed for almost 60 years, becoming the only widely accepted and used equation in the field. Admittedly, Vinen himself expressed concerns whether this equation
may apply to inhomogeneous  situations \cite{Vinen1957}, and in subsequent work attempts have
been made to generalize this equation, but no conclusive results were obtained. To find out more about the history of the problem see the review paper \cite{Nemirovskii2013} and references therein.

The next important step was the work of Schwarz \cite{Schwarz1988} who applied vortex filament method to quantum turbulence to derive an equation for the vortex line density from microscopic principles, but it also underlined the closure problem that results from this approach. Additional and more complicated characteristics of the vortex tangle pop up from the derivation. The two most important
objects that appear naturally are the mean curvature and the anisotropy of the vortex tangle. Accordingly, finding an acceptable equation for $\C L$ for general flows becomes a challenge of finding a reasonable closure in terms
of variables that can be measured to compare observations with theory.

A serious hindrance to the completion of this program was the lack of sufficient data to provide verification of possible theories. This hindrance began to lift recently with the advent of numerical simulations.  Lipniacki \cite{Lipniacki2001} proposed to complement the equation for vortex line density with a coupled equation for the anisotropy parameter in the context of homogeneous but time dependent counterflow turbulence. Further improvement of numerical simulations made it possible to study inhomogeneous counterflow turbulence in a channel  \cite{Khomenko2015,Khomenko2016c,Baggaley2014,Yui2015}. In our own work \cite{Khomenko2015} we suggested a new form of the equation for $\C L$ in a channel which started some discussion in the literature \cite{Nemirovskii16,Khomenko2016c}. The present work is a continuation and an extension of Ref.~\cite{Khomenko2015}. Here  we present a new derivation of an equation for the anisotropy  parameter which generalizes and complements our previous results. In our work we opt to focus on steady inhomogeneous flow, in contrast to Ref.~\cite{Lipniacki2001} which considered homogeneous unsteady flow.  We should state at this point that the derivations
presented below are not mathematically rigorous. They are based on arguments and
estimates. Whenever we can we supplement the analytic arguments with numerical tests.
The state of the art of numerical simulations of quantum fluids is not sufficient
to nail all the assumption made with certainty. So the reader is advised to consider
the present paper as a statement of the state of the art which is possibly not final.
More careful work is needed to reach final conclusions.
Nevertheless the use of numerical simulations allows us to directly check and verify some crucial assumptions; this fundamentally distinguishes our work from some of the previous papers in this field\cite{Jou2011}.

The structure of this paper is as follows: in Sect.~\ref{intro} we review the derivation of the equation of the density of vortex lines and explain how closure bring up the necessity to study new objects like the anisotropy parameter. In Sect.~\ref{aniso} we derive the equation for anisotropy parameter. A discussion
and conclusions are offered in Sect.~\ref{discussion}. The appendix provides
those technical details that are not given explicitly in the text.

\section{\label{intro}equation for the density of vortex lines}

The starting point for the derivation of an equation of motion of the density
of vortex lines is the microscopic equation for a {\em single} vortex line \cite{Schwarz1988}. Denote by
$\bm s(\xi,t)$ a given point in space that belongs to a vortex line which is parameterized by the
arc-length $\xi$. The one writes the equation
\begin{eqnarray} \label{SFVel}
 \frac{d\B s(\xi,t)}{dt} =  \B V_s+ {\bm V}\Sb{BS} (\B s,t)
+(\alpha - \tilde \alpha \bm {s}'\times\big )  \mathbf{s}' \times \B V\sb{ns}(\B s,t)\,  .
 \end{eqnarray}
Here ${\bm V}\Sb{BS} (\B s,t)$ is the velocity generated by the entire vortex tangle, the prime means a derivative along an arc length $d/d\xi$ and $\alpha$ and $\tilde \alpha$ are mutual friction parameters. From this equation we can calculate rate of elongation of the vortex line segment $\delta \xi$:
\begin{equation}\label{segmentlength}
\frac{1}{\delta\xi}\frac{d\delta\xi}{dt}=\bm s'\cdot\frac{d \bm s'}{dt}\ .
\end{equation}
Integrating Eq.~(\ref{segmentlength}) over the vortex tangle provides the change in the density of vortex lines $\C L$. Note that in inhomogeneous condition this wanted equation has the general form:
\begin{equation}
\frac{\partial\C L}{\partial t}+\B \nabla \cdot \C J_{\C L} = \C P_{\C L}-\C D_{\C L}\ .
\end{equation}
where  $\C J_{\C L}$ is the flux of the vortex line density, $\C P_{\C L}$ is the rate of production and $\C D_{\C L}$ the rate of decay of the said density. In the appendix and in Refs.~ \cite{Khomenko2015, Khomenko2016c} it is shown that in a channel geometry where the mean flow is in the $x$-direction and the wall normal direction is $y$ one can derive the approximate equations
\begin{subequations} \label{5}
\begin{eqnarray}
\C J_{\C L} &= & \int \left[\frac{d \bm s}{dt}\right]_y d\xi \approx  -\frac{\alpha}{\kappa} V\sb{ns}\frac{dV\sb s}{dy}\ ,\label{J}\\
\C P_{\C L} &\approx & \alpha\bm V\sb{ns}\cdot \bm I_{l0} \<\varkappa\>\C L\ , \\
\C D_{\C L} &= & \alpha\beta\<\varkappa^2\>\C L \ .
\label{EqL}
\end{eqnarray}
Here $\varkappa\equiv|s''|$ is the curvature of the vortex line  and the dimensionless vector $\bm I_{l0}$ is defined by:
\begin{equation}
\bm I_{l0}=\<\bm b\>, \, \bm b= \bm s'\times \bm s''/\varkappa\ .
\end{equation}\end{subequations}

The derivation of these equations is based on the following assumptions:
\begin{enumerate}
  \item Separation of scales; we assume that the ``macroscopic" variable $V\sb{ns}$ changes slowly in space in comparison to rate of spatial change  of $\bm s'$ and $\bm b$. Accordingly the average of their product can be estimated as a product of the averages.
  \item  We neglect the term in Eq.~(\ref{SFVel}) proportional to $\tilde \alpha$. The reason is that in counterflow conditions this coefficient is much smaller than $\alpha$.
  \item  We assume that the derivative $dV\sb{ns}/d\xi$ along vortex lines is negligible on the average.
\end{enumerate}
 Equations\,\eqref{5} expose the appearance of new fields, i.e. the objects $\<\varkappa\>,\<\varkappa^2\> $ and $\bm I_{l0}$. To be able to close the system of equations these objects should be either expressed in terms of known variables ($\C L$, $V\sb{ns}$, ...) or supplemented by equations for the new objects.

To proceed, another fundamental assumption is called for. We assert that there exists only one typical length-scale in the problem which is inter-vortex distance. Therefore the radius of curvature should be proportional to it. This is a closure relation that reads $\<\varkappa\> \simeq c_1 \C L^{1/2}$. This assumption is expected to be reasonable for relatively low gradients of counterflow velocity. We tested it and found that it works well for  parabolic profiles of the normal velocity in the case of T1 turbulence in a channel, cf. Ref. \cite{Khomenko2015}.

The closure of $I_{l0}$ is more complicated. In Ref. \cite{Khomenko2015}, in analogy with the Vinen equation,  we assume that $I_{l0}$ can be expressed in terms of $V\sb{ns}$ and $\C L$ only. Being dimensionless, it must be a function of the dimensionless variable $\zeta=V\sb{ns}/(\kappa\C L^{1/2})$. Based on numerical experiments we concluded  that taking $I_{l0}\propto\zeta^2$ gives a good match for all the available data obtained by simulations in a channel. Below we  generalize this to other geometries.
With these explicit assumptions one obtains the closure  for Eq.~(\ref{EqL}):
\begin{subequations}\label{6}
 \begin{eqnarray}
\C J_{\C L} &\approx & -\frac{\alpha}{\kappa} V\sb{ns}\frac{dV\sb s}{dy}\,, \\
\C P_{\C L} &\approx & \alpha C\sb{prod} V\sb{ns}^3\sqrt{\C L}/\kappa^2 \,, \\
\C D_{\C L} &\approx & \alpha\beta C\sb{dec} \C L^2\ .
\end{eqnarray}\end{subequations}

\begin{figure}
  \includegraphics[scale=0.35 ]{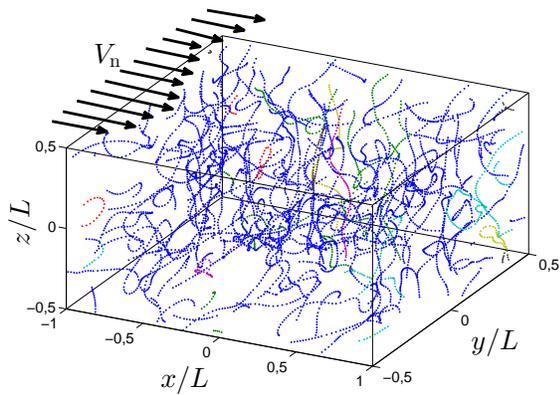}
 \caption{ \label{F:1} Sketch of the computational setup.}

\end{figure}
\begin{figure}
   \includegraphics[scale=0.45 ]{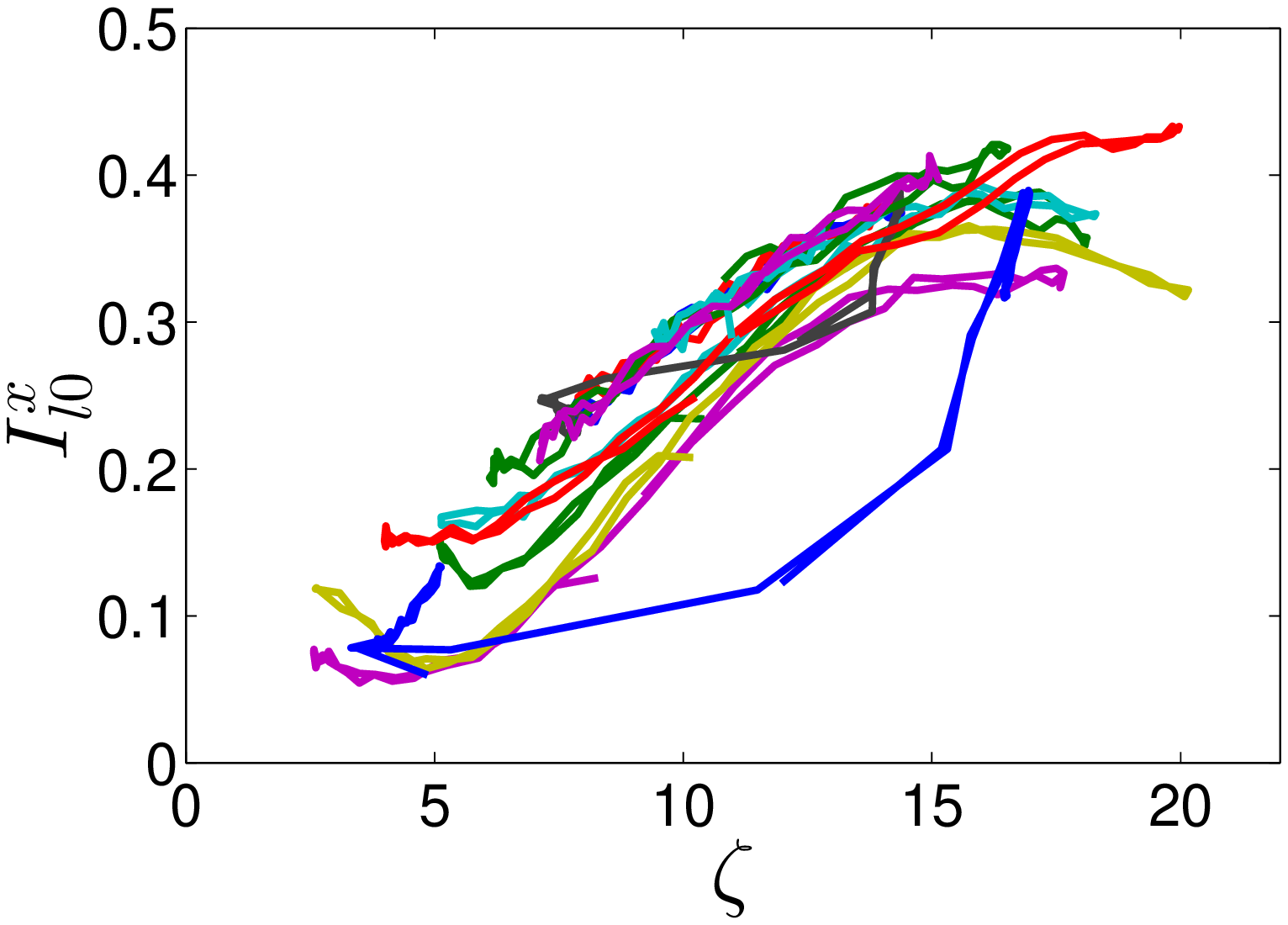}
   \includegraphics[scale=0.45 ]{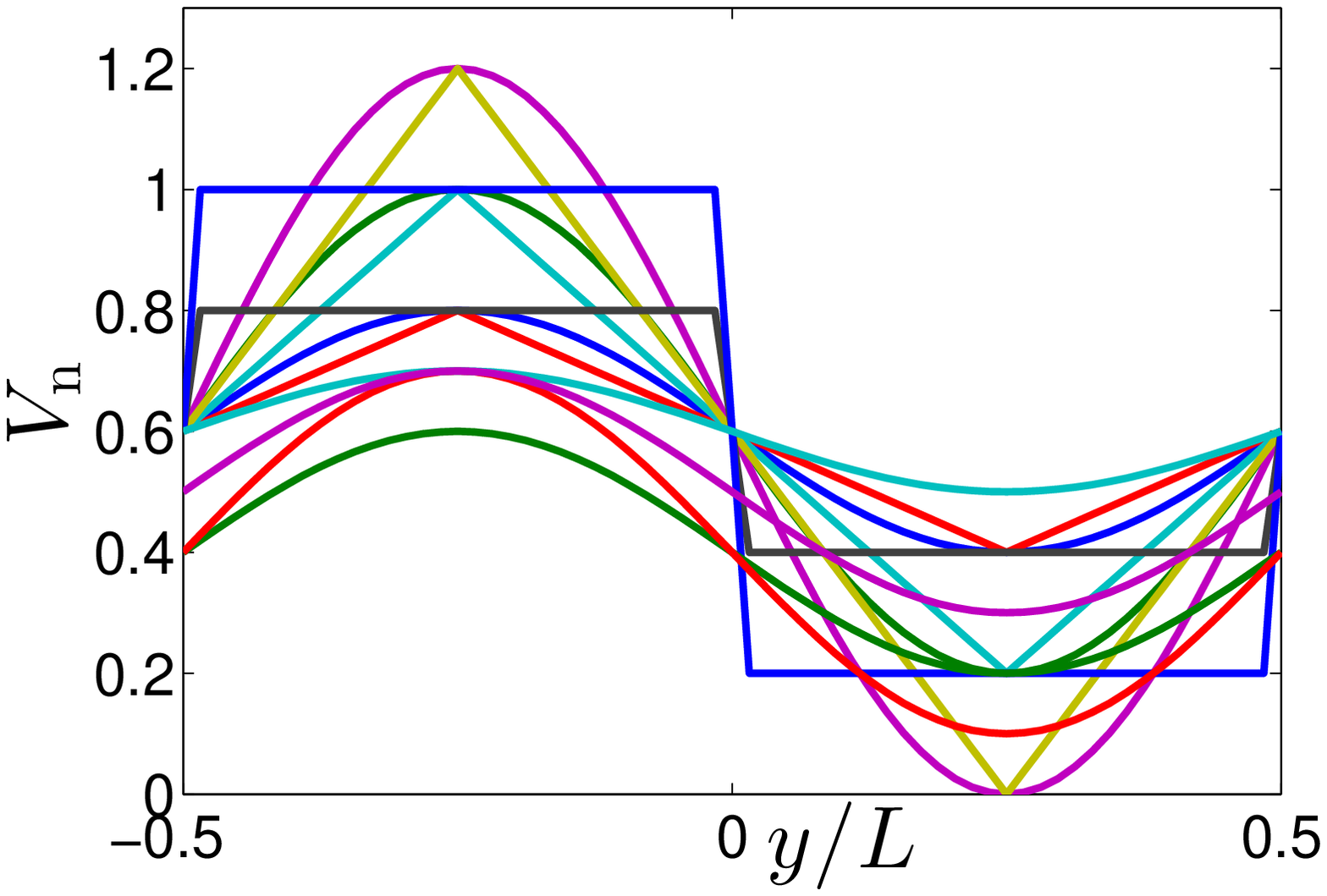}
 \caption{ \label{F:2} Upper panel: examples of imposed normal velocity profiles in the simulation channel. Lower panel: parametric plot  $[\zeta(y),I^x_{l0}(y)]$ of the stream-wise projection of  the anisotropy parameter.}

\end{figure}

\section{Equation for the anisotropy parameter }
\label{aniso}
\begin{figure}
   \includegraphics[scale=0.35 ]{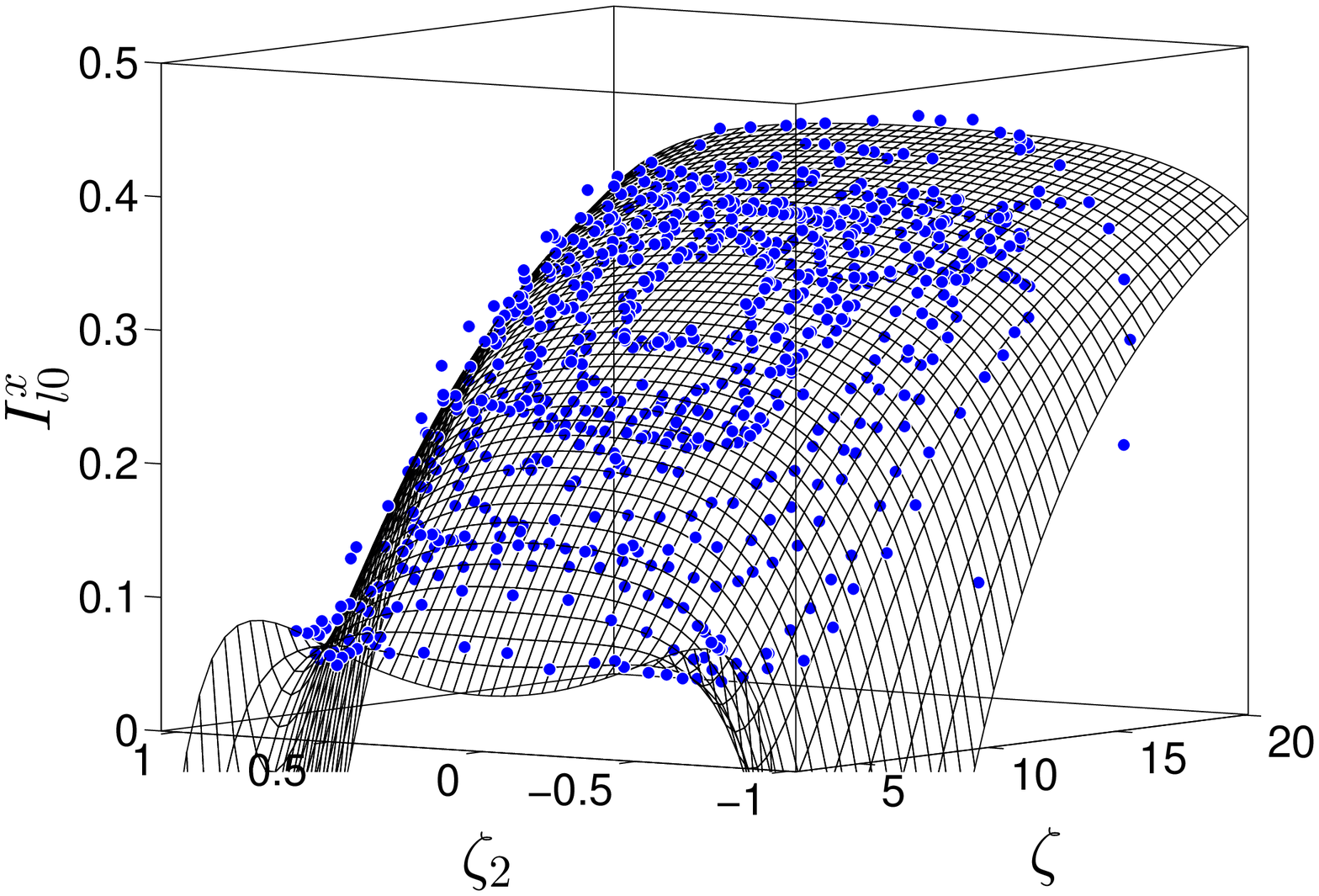}
  \includegraphics[scale=0.35 ]{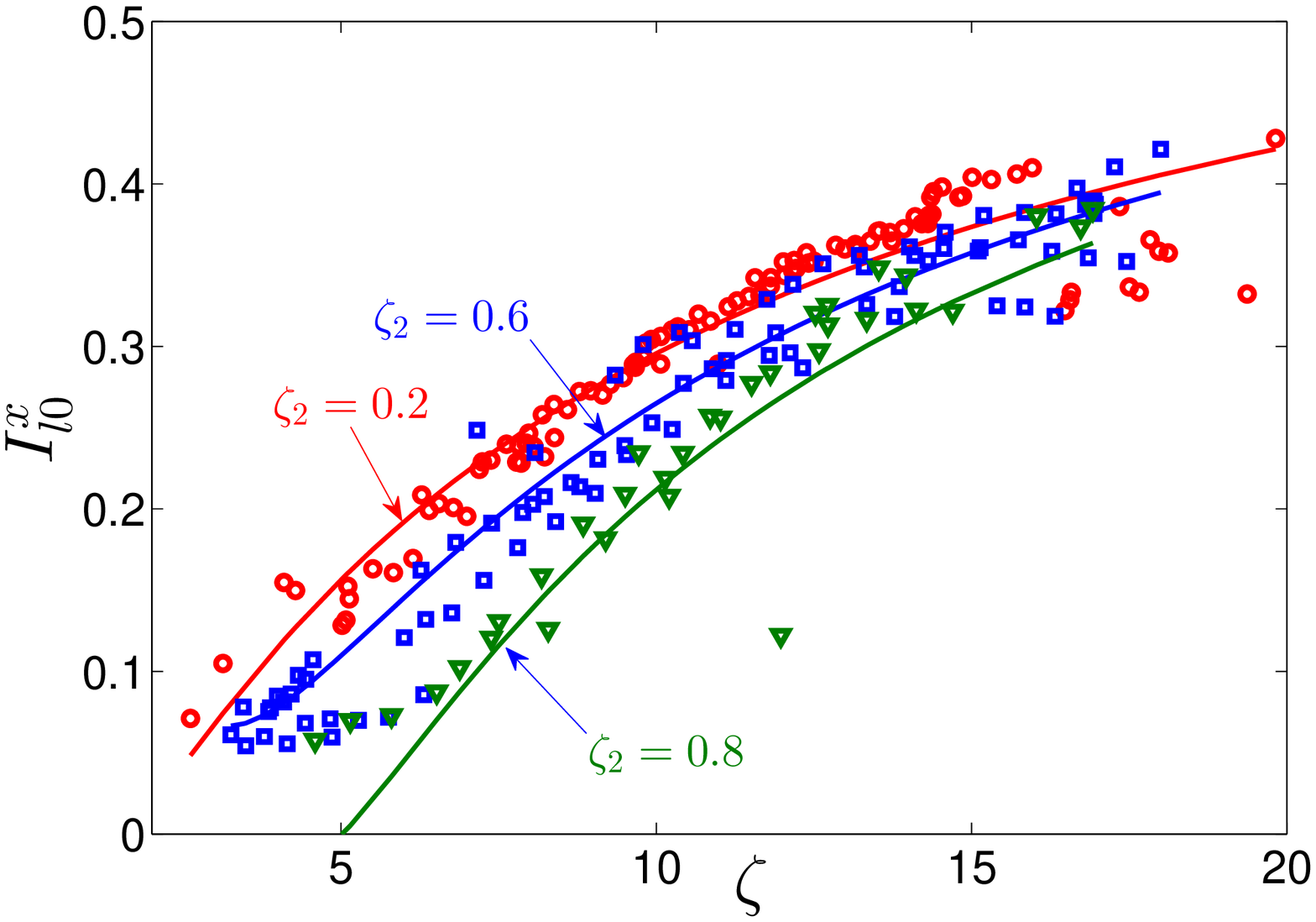}
\caption{\label{F:3} Parametric plot of $[\zeta(y),\zeta_2(y),I_{l0}(y)]$ for all the normal velocity profiles shown in \Fig{F:2}. Upper Panel: The surface embedded in 3-dimensions which spans all the available data points. Lower Panel: $I_{l0}(y)$ as a function of $\zeta$ for different values of $\zeta_2$, as marked in the plot. Points are data from all the simulations for $\zeta_2=0.2\pm 0.1$ (red), $0.6\pm 0.1$ (blue), $0.8\pm 0.1$ (green). Lines of corresponding colors denote the fit [cf the surface shown in the Upper Panel].
}

\end{figure}

Admittedly, while the closure
 for $I_{l0}$ matches well with the available numerical results for T1 counterflow turbulence in  the channel, its derivation did not provide adequate physical intuition or an explanation why it gave
a good agreement with the data. Accordingly it is not clear how general this closure is.
To test the limits of its applicability we did a series of additional numerical simulations with various different profiles of the normal velocity shown in the upper panel of \Fig{F:1}. The simulations were performed using the full Biot-Savart calculation, employing vortex filament methods as explained in detail in Ref.~\cite{Kondaurova2014}. Using periodic boundary conditions , we have a greater freedom of choice for the normal velocity profile. The simulations were carried out in a computational box of size $2L\times L\times L, L=0.1$ cm with 3-periodic boundary conditions (cf. Fig\ref{F:1}) for $T=1.6\,$K and $V\sb n=(0.4\div 1.0)$ cm/s.  In all these simulations we measured the anisotropy parameter $I_{l0}$ and plotted the results in the lower panel of \Fig{F:2} as a parametric plot $[I_{l0}(y),\zeta(y)]$. The results convey the clear message that $I_{l0}$ is not a function of $\zeta$ only. Additional dimensionless variables in addition to $\zeta$ appear to be necessary.
Candidates are for example
\begin{eqnarray}
\zeta_2=\frac{1}{\kappa \C L}\frac{dV_s}{dy}\,, \quad   \zeta_3=\frac{1}{\kappa \C L^{3/2}}\frac{d^2V_s}{dy^2}\,,  \dots
\end{eqnarray}
As a test of this conclusion we show in \Fig{F:3} the results
of allowing the anisotropy parameter to depend on two
variables, i.e. $\zeta$ and $\zeta_2$.
An improvement in the data representation is evident. To find the two-variable parametric surface, we used general polynomial and Pad\'e approximants. The resulting  mean square deviation was reduced by 50 \% compared to the fit that depended on $\zeta$ only. Accordingly we turn
now to a derivation based on microscopic relations.

\subsection{Derivation based on microscopic relations}

To find an expression for the anisotropy parameter $I_{l0}$ in a systematic way we will derive its equation of motion. We start again from \Eq{SFVel}. In the appendix we show that applying this equation to the unit vector $\bm b$ we get:
\begin{eqnarray}\label{Eqb}
\frac{d \bm b}{dt}=\alpha\varkappa \bm V_{\rm ns}^0-\alpha \varkappa \left(\bm b \cdot \bm V_{\rm ns}^0\right)\bm b + \beta \tau^2 \bm s'' /\varkappa \ ,
\end{eqnarray}
where $\tau$ is a torsion of the vortex line. This equation can be rewritten in the following form:
\begin{equation}
\frac{d \bm b}{dt}=\alpha \varkappa \bm b \times \left(\bm V_{\rm ns}^0 \times \bm b  \right) + \beta \tau^2 \bm s'' /\varkappa \ .
\label{sec}
\end{equation}
Note that \Eq{sec} is very similar to the equation for vortex rings that was used in Ref.~\cite{Lipniacki2001} for $I_{l0}$ in a homogeneous flow. In an inhomogeneous and anisotropic flow the equation satisfied by $I_{l0}$
will again have a production term denoted as $\C P_{I_{l0}}$, a decay term, denoted as $\C D_{I_{l0}}$, and a flux term denoted as $\bm {{\hat{\C J}}_{I_{l0}}}$. Two of these can be obtained easily.
Integrating \Eq{Eqb} over the tangle provides the production term for $I_{l0}$. The flux term can also
be found analytically,
\begin{equation}
\bm {{\hat{\C J}}_{I_{l0}}}=\Big\langle\frac{d \bm s}{dt}\otimes\bm b\Big\rangle\ .
\end{equation}
\begin{figure}
    \includegraphics[scale=0.45 ]{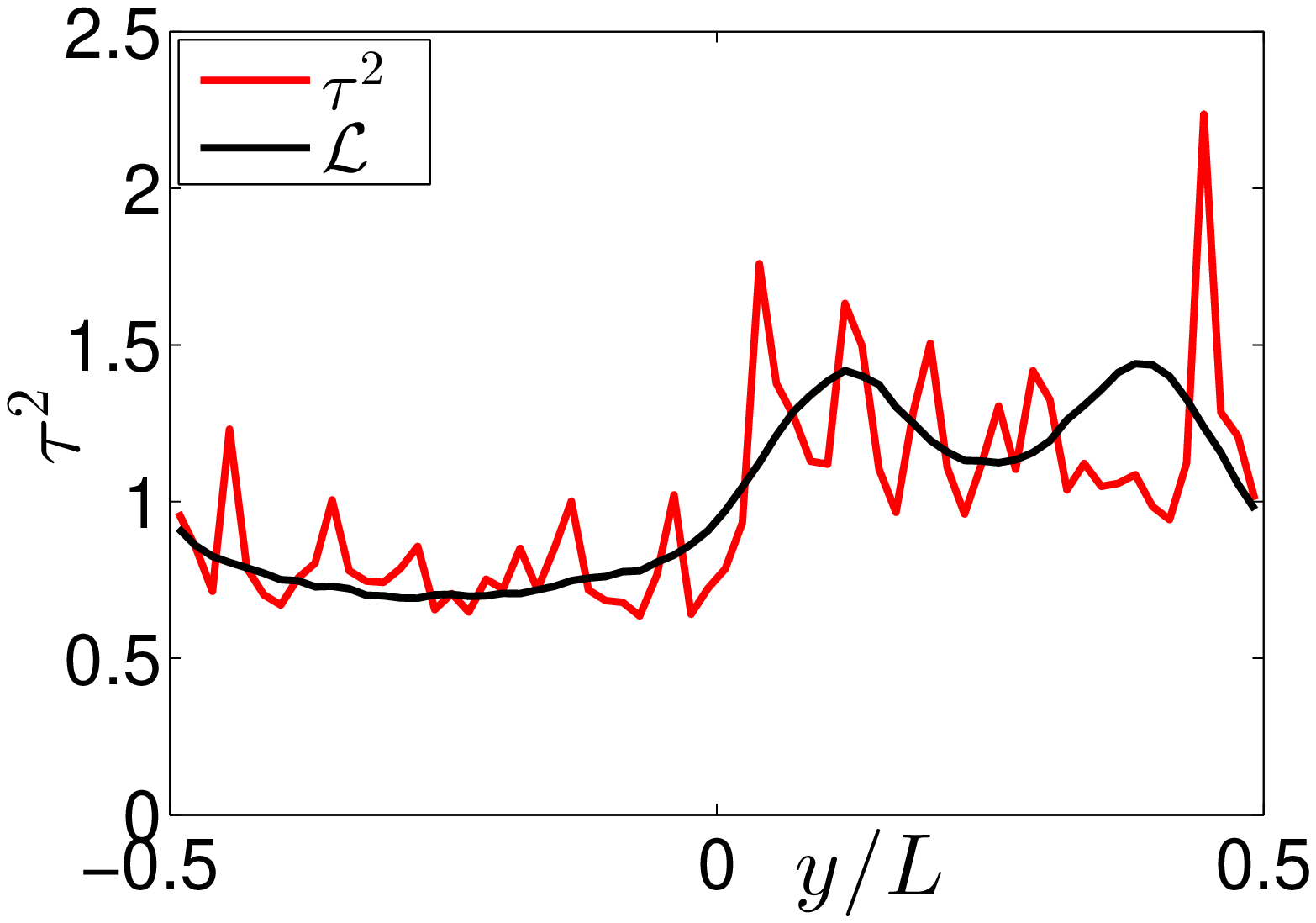}
   \includegraphics[scale=0.45 ]{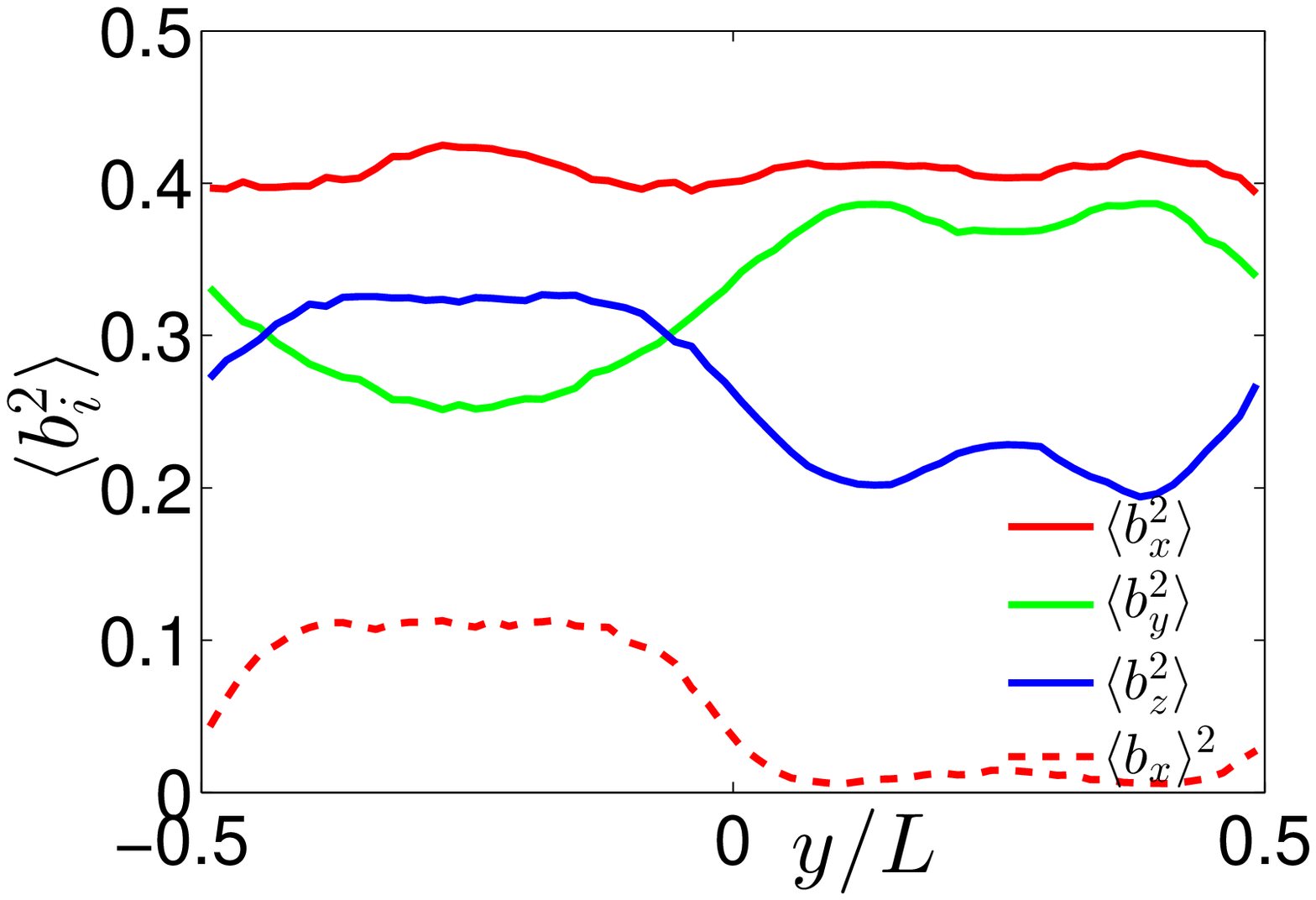}
\caption{ \label{F:4} Upper panel: the profiles of $\tau^2(y)$ and the normalized density of vortex lines $\C L(y)/\langle \C L \rangle$,  plotted for the sine-like normal velocity profile. Lower panel:  the corresponding profiles of $\<b_i^2\>$ and $\<b_i\>^2$.}

   \end{figure}

The last ingredient, the decay, is governed by the effect of vortex reconnection. This term is difficult to derive from the microscopic equation and at this point it can only be written down phenomenologically. The effect of vortex reconnection is to locally form sharp  tips \cite{Buttke98,TYB11,ZCBB12,Boue2013} that can be oriented in any spatial direction which is determined by the relative  orientation of the two reconnecting vortex lines.  This tends to destroy any pre-orientation of the vortex lines. Based on this picture we can assume that in each event of reconnection a region which is affected by it will loose its anisotropy $I_{l0}$. Phenomenologically we can write the decay due to reconnections in the following way:
\begin{equation}
\C D_{I_{l0}}\approx\frac{C}{\<\varkappa\>\C L}\frac{dN\sb{rec}}{dt}I_{l0}\ .
\end{equation}
Here $dN\sb{rec}/dt$ is the reconnection rate, $1/\varkappa$ is the radius of curvature which defines the region affected by the reconnection event and $I_{l0}$ is the value of anisotropy that was lost during the reconnection.

Now that all the necessary objects are  expressed in terms of macroscopic fields, we recognize that ones more there appeared new quantities, i.e.  $\langle b_x^2 \rangle$, $\tau^2$ and $ dN\sb{rec}/dt$ that should be modeled.
We will start with torsion. Torsion is a second curvature and we expect that the radius of torsion would be proportional to the inter-vortex distance:

\begin{equation}
\<\tau^2\> \approx C_{\tau} \C L \ .
\end{equation}
Note that this assumption is similar  to the previously performed closure for the curvature $\varkappa$  and is
expected to have  similar validity. Unfortunately at present we can not test this assumption numerically due to the insufficient accuracy of our simulations. Nevertheless preliminary results shown in the upper panel of \Fig{F:3} look promising. Regarding the object $\<b_x^2\>$ there are two limiting cases that could be considered: (i) strongly polarized tangle: in this case $\<b_x^2\>\approx\<b_x\>^2$; (ii) a fully isotropic case: $\<b_x^2\>\approx 1/3$.  Based on the results of many numerical simulation we can say that in the case of counterflow turbulence (T1 regime) we are closer to the second case, and even though $\<b_x^2\>$ is slightly larger than $(\<b_y^2\>+\<b_y^2\>)/2$, it can be considered as a constant with reasonable accuracy,
\begin{equation}
\<b_x^2\> \approx \mbox{const.}
\end{equation}
Evidence is provided in the lower panel of \Fig{F:4}.
\begin{figure}
 \hskip -0.4cm  \includegraphics[scale=0.45 ]{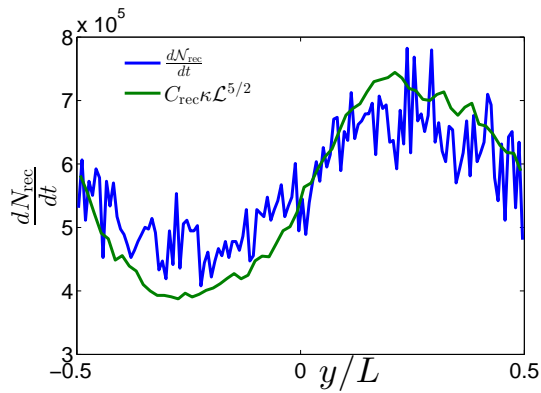}
\caption{\label{F:5} The profile of the reconnection rate and its estimate via the closure
discussed in the text. In this example the imposed normal velocity profile is sine-like.}
\end{figure}

The last object is the vortex reconnection rate. Motivated by dimensional reasoning and supported by results from homogeneous turbulence \cite{Barenghi2004,Nemirovskii2006}, we can model the reconnection rate profile by:
 \begin{equation}
\frac{dN\sb{rec}}{dt} \approx C\sb{rec}\kappa \C L^{5/2} \ .
\end{equation}
The numerical data presented in \Fig{F:5} support this conclusion and see also the results in Ref.\,\cite{Baggaley2014}.

\subsection{Putting things together}

Summarizing together all the discussed results we end up with
a system of equations:
\begin{subequations}\label{15}
\begin{equation}
\frac{\partial I_{l0}}{\partial t}+\frac{\partial \C J_{I_{l0}}}{\partial y} =\C P_{I_{l0}}-\C D_{I_{l0}}\,,
\end{equation}
\begin{eqnarray}\label{IlEq}
\C J_{I_{l0}} &\approx & \frac{1}{\C L}\Big(-\frac{\alpha}{\kappa} V\sb{ns}\frac{dV\sb s}{dy}I_{l0}+C\beta\frac{dV_s}{dy}\Big)\,, \\
\C P_{I_{l0}} &\approx & \alpha  C_1  V\sb{ns} \C L^{1/2} +C_2\frac{\beta}{\C L} \frac{d^2 V_s}{dy^2}\,, \\
\C D_{I_{l0}} &\approx &  C\sb{dec} \kappa \C L I_{l0}\ .
\end{eqnarray}
\end{subequations}
Note that in \Eq{IlEq} we used results from Ref.~\cite{Khomenko2016b} to model $\<s''\>$.

\section{Discussion and conclusions}
\label{discussion}
The central result of this paper is the equation of motion for anisotropy parameter $I_{l0}$; we showed that it is possible to close it in terms of known variables. Together with the equation for the density of vortex lines we possess now a set of equations that allows a calculation of the profile of the anisotropy parameter and the vortex line density self-consistently, relaxing the assumption on the production term made in Ref.\cite{Khomenko2015}. The approach described above has pros and cons. On the negative side, we recognize that a number of new assumptions
were made whose verification is currently  beyond our computational capabilities. Of particular concern is the phenomenological model for the anisotropy decay due to reconnections, and the precise modeling of the torsion and the flux of $I_{l0}$. On the positive side the work described above prepares a theoretical background for future numerical simulations.  It underlines the missing links in our current understanding of the subject. Moreover, even on the basis of what is known now we have enough information from numerical simulations to afford making  simplifications that are relevant for the study of counterflow turbulence. Finally, possible future experiments in inhomogeneous and anisotropic superfluid
flows should also be carefully designed and analyzed to shed further light on the issues discussed above. Indeed, a proper description of the physics of vortex lines requires knowledge of both the vortex line density and the anisotropy of the vortex tangle;  the measurement of at least two different fields is needed. For example, one can use the fact that the attenuation of second-sound depends on the angle between its propagation direction and the vortex line direction \cite{Babuin12}.
Thus, measuring the attenuation along two orthogonal directions one can extract the total line density
as well as information about the the anisotropy of the vortex tangle. Probably even more informative  would be experiments of a counterflow and pure superflow in a rectangular channel with a large aspect ratio  measuring the second sound attenuation, propagating in all three directions.  This type of experiments, possibly combined  with visualization techniques, may expand our understanding of counterflows in superfluid $^4$He, and in particular, shed light on the importance of the  vortex tangle anisotropy.

\acknowledgements

IP is grateful for the hospitality of Prof. Poul Henrik Damgaard at the Neils Bohr International
Academy.

\appendix
\section{Microscopic Derivations}
\subsection{Equation for the density of vortex lines.}

We will start from \Eqs{SFVel} and \Eq{segmentlength}:
\begin{equation}\label{gen}
\frac{1}{\delta \xi}\frac{d\delta \xi }{dt} =  \alpha \bm V\sb{ns}(\B s,t)\cdot (\B s'\times \B s'')+\B s'\cdot {\B V\sb{nl}^s}'-\tilde \alpha \B s''\cdot \B V\sb{ns} \ .
\end{equation}
The first simplification that we make is neglecting the term proportional to $\tilde \alpha$. This simplification is justifiable for temperatures higher than T=1.3 K when $\tilde \alpha$ much smaller than $\alpha$, but might be reconsidered for lower temperatures.
Second, we can argue that the second term on the RHS of \Eq{gen} is in fact negligible. To see this note that $\B V\sb{nl}$ is a macroscopic field that changes slowly on the length-scale of inter-vortex distance. For the present flow which is  inhomogeneous in the $y$-direction, we note that due to symmetry all the average values depend only on the y-coordinate,  and their only non-zero component is in the x-direction. In contrast,  $\<s'_y\>=\<s'_x\>=0$. Together this is a strong reason to assert that this term is negligible compared to the first term in Eq.~(\ref{gen}). Indeed  in our numerical simulations we could confirm this assertion, cf.  Ref.~\cite{Khomenko2016c}.

Next we can decompose the counterflow velocity $V\sb{ns}$ into a local $-\beta \bm s'\times \bm s''$ and slowly-varying nonlocal $V\sb{ns}^0$ components:
\begin{equation}
\frac{1}{\delta \xi}\frac{d\delta \xi }{dt} \approx  \alpha V\sb{ns}^0\cdot (\B s'\times \B s'')- \alpha \beta |s''|^2 \ .
\end{equation}
Integrating this equation over the tangle results in an equation for the production and decay of the vortex line density:
\begin{eqnarray}
\C P_{\C L}&=& \frac{\alpha}{\Omega}\int \B V\sb{ns}^0\cdot (\B s'\times \B s'')d\xi\approx \alpha \C L \<\varkappa\>V\sb{ns}\cdot \<b\>\,, ~~~~~~\\
\C D_{\C L}&=&\frac{1}{\Omega}\int\alpha \beta |s''|^2 d\xi \approx \alpha \beta \C L \<\varkappa^2\>\ .
\end{eqnarray}

\subsection{The flux of the vortex lines density }

The flux of the density of vortex lines is defined by:
\begin{equation}
\BC J_{\C L }=\Big\langle \frac{d\bm{s}}{dt} \Big\rangle\ .
\end{equation}
In the considered geometry only the $y$-component survives:
\begin{equation}
[\C J_{\C L}]_y\approx\left\langle \alpha s'_z V\sb{ns} \right\rangle\approx  \alpha \left\langle  s'_z \right\rangle  V\sb{ns}=-\frac{\alpha}{\kappa} V\sb{ns}\frac{dV\sb s}{dy}\ .
\end{equation}
Note that in general the vortex flux can also contain  other terms \cite{Khomenko2016c}, for example a diffusive component \cite{Tsubota2003,Nemirovskii2010}; however if the tangle is polarised, as in the case of counterflow or rotating turbulence, these terms are expected to be negligible. Nevertheless, in particular geometries\cite{Saluto16} they may become important.
\subsection{The equation for the anisotropy parameter}
We start from Eq.~(\ref{SFVel}) after discarding the term proportional to $\tilde \alpha$:
\begin{equation}
\frac{d\bm s}{dt}=\bm V_s+\alpha \bm s'\times \bm V_{\rm ns}.
\end{equation}
The derivative for binormal vector $\bm s'\times \bm s''$ reads
\begin{eqnarray}
\frac{d}{dt}(\bm s'\times \bm s'')=\frac{d\bm s'}{dt}\times \bm s''+ \bm s' \times \frac{d \bm s''}{dt} \ .
\end{eqnarray}
Using these equations together gives:
\begin{eqnarray}\label{eqIl0_0}\nonumber
&&\frac{d \bm }{dt}(\bm s'\times \bm s'')=\bm V_s' \times \bm s'' + \alpha (\bm s''\times \bm V_{\rm ns})\times \bm s''\\ \nonumber
&&+\alpha (\bm s'\times \bm V_{\rm ns}')\times \bm s''+\bm s'\times \bm V_{s}''+\alpha \bm s'\times (\bm s'''\times \bm V_{\rm ns})\\
&&+2\alpha \bm s'\times (\bm s''\times \bm V_{\rm ns}')+\alpha \bm s'\times (\bm s'\times \bm V_{\rm ns}'').
\end{eqnarray}
To express higher derivatives in terms of $\bm s'$ and $\bm s''$ we can use the Frenet-Serret formula \cite{wiki}:
\begin{equation}
\bm s'''=-\varkappa^2\bm s'+ \tau \bm s' \times \bm s'' +\frac{\varkappa'}{\varkappa} \bm s'' \ ,
\end{equation}
where $\varkappa$  is the curvature and $\tau$ is the torsion.
At this point it is advantageous to split $V\sb{ns}$ into a local term $-\beta \bm s' \times \bm s''$ and a non local term $ V\sb{ns}^0$:
\begin{equation}
\bm V\sb{ns}=\bm V\sb{ns}^0-\beta \bm s' \times \bm s'' \ .
\end{equation}
If we neglect terms with derivatives of the nonlocal velocity component, we get:
\begin{eqnarray}
\nonumber
&&\frac{d}{dt}(\bm s' \times \bm s'')=2\alpha\varkappa^2\bm V\sb{ns}^0+\alpha\bm s'\left[ -\beta \varkappa^2 \tau  -\varkappa^2 (\bm s'\cdot \bm V\sb{ns}^0) \right] \\
&&+\bm s'' \left[ \beta \tau^2 -\alpha (\bm s'' \cdot \bm V\sb{ns}^0) \right]-\alpha \beta \left[ 2\varkappa^2 +\tau^2 \right] (\bm s' \times \bm s'')
\end{eqnarray}

The next step of simplification is to neglect terms proportional to $\tau$. The torsion, in contrast to the curvature, is not positive definite and we can assume that the mean value of $\tau$ is negligible. Our current numerical simulations are not sufficiently  precise to test this assertion, although preliminary results indicate that this assumption is correct.

Now having an equation for $\bm s' \times \bm s''$ we can write an equation for the unit vector $\bm b$:
\begin{equation}
\frac{d \bm b}{dt}=\frac{1}{\varkappa}\left(\frac{d\bm s' \times \bm s''}{dt} -\bm b \frac{d \bm s' \times \bm s''}{dt} \cdot \bm b\right) \ .
\end{equation}
Substituting $d(\bm s' \times \bm s'')/dt$:
\begin{eqnarray}
\frac{d \bm b}{dt}&=&\alpha\varkappa\bm V_{\rm ns}^0-\alpha \varkappa \left(\bm b \cdot \bm V_{\rm ns}^0\right)\bm b +\beta \tau^2 \bm s''/\varkappa
\end{eqnarray}

\subsection{Flux of the anisotropy parameter}
The flux of the anisotropy parameter $I_{l0}$ can be written as:
\begin{equation}
 {\hat{\BC J}}_{\B I_{l0}}=\Big\langle \frac{d \bm s}{dt} \otimes  \bm b\Big\rangle\ .
\end{equation}
In the channel geometry  only the  $xy$ component of the tensor survives:
\begin{equation}
 [{\C J}_{I_{l0x}}]_y=\Big\langle \Big[\frac{d \bm s}{dt}\Big]_y  b_x \Big\rangle\approx \langle \beta \varkappa  b_x b_y \rangle+ \langle \alpha s'_z V\sb{ns}  b_x\rangle \ .
\end{equation}
The first term is given by local velocity and with a good accuracy can be expressed as:
\begin{eqnarray} \nonumber
\<b_x b_y\>&=&\<\frac{s'_y s'_z s''_z s''_x-s'_x s'_y {s''_z}^2-{s'_z}^2 s''_z s''_x+s'_x s'_z s''_y s''_z}{\varkappa^2}\>  \\
&\approx & -\<s_x's_y'\>/3\,,
\end{eqnarray}
while for second term we make the uncontrolled approximation:
\begin{eqnarray} \nonumber
\langle \alpha s'_z V\sb{ns}  b_x\rangle \approx \< \alpha s'_z V\sb{ns}^0 \>\<  b_x\>- \alpha \beta \< s_z'\>\< \varkappa b_x^2\>\ .
\end{eqnarray}
This approximation needs to be verified in future numerical simulations.


\begin{thebibliography}{99}
\bibitem{tabeling98}J. Maurer  and P. Tabeling,\emph{Local investigation of superfluid turbulence}, Europhys.  Lett. , \textbf{43}, 29 (1998)
\bibitem{Vinen1957}W.F. Vinen. \emph{Mutual friction in a heat current in liquid helium II. III. Theory of the mutual friction}. Proc. R. Soc. Lond. A \textbf{242}, 493 (1957).
\bibitem{Nemirovskii2013}S. Nemirovskii, \emph{Quantum turbulence: Theoretical and numerical problems}. Physics Reports, \textbf{524}, 85 (2013).
\bibitem{Schwarz1988}K. W. Schwarz, \emph{Three-dimensional vortex dynamics in superfuid $^4$He: Homogeneous superfluid turbulence}. Phys. Rev. B, \textbf{38}, 2398 (1988).
\bibitem{Lipniacki2001}T. Lipniacki. \emph{Evolution of the line-length density and anisotropy of quantum tangle in He4}. Phys. Rev. B, \textbf{64} 214516 (2001).
\bibitem{Khomenko2015}D. Khomenko, L. Kondaurova, V.S. L'vov, P.  Mishra, A. Pomyalov, I. Procaccia. \emph{Dynamics of the density of quantized vortex lines in superfluid turbulence}.  Phys. Rev. B. \textbf{91},80504(R) (2015).
\bibitem{Nemirovskii16}S. K. Nemirovskii, \emph{Comment on “Dynamics of the density of quantized vortex lines in superfluid turbulence”}. Phys. Rev. B \textbf{94}, 146501 (2016).
\bibitem{Khomenko2016c}D. Khomenko, V.S. L’vov, A. Pomyalov, and I. Procaccia. \emph{ Reply to ”comment on ’Dynamics of the density of
quantized vortex lines in superfluid turbulence’ ”}. Phys. Rev. B ,\textbf{942}, 146502 (2016).
\bibitem{Baggaley2014}A. W. Baggaley and J. Laurie. \emph{Thermal Counterflow in a Periodic Channel with Solid Boundaries}. JLTP, \textbf{178}, 35 (2014).
\bibitem{Yui2015}S. Yui and M. Tsubota. \emph{Counterflow quantum turbulence of He-II in a square channel: Numerical analysis with nonuniform flows of the normal fluid}. Phys. Rev. B \textbf{91}, 184504 (2015).
\bibitem{Jou2011}D. Jou, M. Mongiov\'{i}, M. Sciacca,  \emph{Hydrodynamic equations of anisotropic, polarized and inhomogeneous superfluid vortex tangles}. Physica D, \textbf{40}, 249 (2011).
\bibitem{Kondaurova2014}L. Kondaurova, V.S. L'vov, A. Pomyalov, I. Procaccia, \emph{Structure of a quantum vortex tangle in He-4 counterflow turbulence}.  Phys. Rev. B. \textbf{89}, 014502 (2014).
\bibitem{Buttke98}T.F. Buttke, \emph{Numerical study of superfluid turbulence in the selfinduction approximation}, J. Comput. Phys. \textbf{76}, 301 (1988).
\bibitem{ZCBB12}S. Zuccher, M. Caliari, A. W. Baggaley, and C. F. Barenghi,  \emph{Quantum vortex reconnections}. Phys. Fluids \textbf{24}, 125108 (2012).
\bibitem{TYB11}R. Tebbs, A.J. Youd, C.F. Barenghi, \emph{The Approach to Vortex Reconnection}, J. Low Temp Phys  \textbf{162}, 314(2011).
\bibitem{Boue2013}   L. Bou´e, D. Khomenko, V.S. L'vov, and I. Procaccia. \emph{Analytic solution of the approach of quantum vortices towards reconnection}. Phys. Rev. Lett., \textbf{111}, 145302 (2013).
\bibitem{Barenghi2004} C. F. Barenghi and D. C. Samuels. \emph{Scaling laws of Vortex Reconnections}. JLTP, \textbf{136}, 281 (2004).
  \bibitem{Nemirovskii2006}  S. K. Nemirovskii. \emph{Evolution of a Network of Vortex Loops in He- II : Exact Solution of the Rate Equation}. Phys. Rev. Lett., \textbf{96}, 015301 (2006).
  \bibitem{Khomenko2016b}D. Khomenko, V. S. L'vov, A. Pomyalov, and I. Procaccia. \emph{Mechanical momentum transfer in wall-bounded superfluid
turbulence}. Phys. Rev. B, \textbf{93}, 134504 (2016).
 \bibitem{wiki}Wikipedia  \emph{Frenet-Serret formulas}, https://en.wikipedia.org/wiki/Frenet
\bibitem{Tsubota2003}M. Tsubota, T.Araki, W. Vinen. \emph{Diffusion of an inhomogeneous vortex tangle}. Physica B, \textbf{329-333}, 224 (2003).
\bibitem{Nemirovskii2010}S. Nemirovskii. \emph{Diffusion of inhomogeneous vortex tangle and decay of superfluid turbulence}. Phys. Rev. B, \textbf{81}, 64512 (2010).
\bibitem{Babuin12} S. Babuin,  M. Stammeier,  E. Varga,  M. Rotter  and L. Skrbek, \emph{Quantum turbulence of bellows-driven 4He superflow: Steady state},
Phys. Rev. B \textbf{86}, 134515 (2012).
\bibitem{Saluto16} L. Saluto, M. S. Mongiov\'{i}, \emph{Inhomogeneous vortex tangles in counterlow superfluid turbulence: flow in convergent channels}, Commun. Appl. Ind. Math. \textbf{7}, 130–149(2016).

\end{thebibliography}

\end{document}